# Vacuum in Accelerators for Mechanical & Materials Engineering

*V. Baglin*
CERN, Geneva, Switzerland

**Abstract**
Vacuum systems are an intrinsic part of any accelerator around the world: all particles circulate under vacuum. This lecture gives rudiments on the fundamentals of vacuum science such as units, ideal gas law, partial pressure, mean free path, flow of molecules, conductance, pumping speed, and outgassing. An overview of standard vacuum instruments for pressure measurement and pumping is presented. Finally, the specificities of accelerator vacuum systems are introduced, discussing some design, construction, installation, and commissioning aspects, as well as fundamentals of beam-vacuum interactions with synchrotron radiation and electron cloud.

Preamble

Readers interested in Vacuum Science and Technology will find a wealth of information in various books, particularly in the excellent CERN Accelerator School proceedings, authored by my esteemed CERN colleagues from whom I have gained invaluable knowledge [1, 2, 3, 4, 5, 6, 7, 8].

## 1 Vacuum fundamentals

The pressure is defined as the force exerted by the molecules on the vacuum chamber wall per unit area.

The pascal, Pa, is the Standard International unit, 1 Pa = 1 N/m$^2$. However, in Vacuum Science and Technology, other units are routinely used such as mbar in Europe and Torr in US (1 Torr = 1 mm of Hg in Torricelli's tube). Table 1 provides some units and their conversions to others.

**Table 1:** Conversion table between usual pressure units

| Converts to | Pa | kg/cm$^2$ | Torr | mbar | bar | atm |
| --- | --- | --- | --- | --- | --- | --- |
| 1 Pa | 1 | 10.2×10$^{-6}$ | 7.5×10$^{-3}$ | 10$^{-2}$ | 10$^{-5}$ | 9.81×10$^{-6}$ |
| 1 kg/cm$^2$ | 98.1×10$^3$ | 1 | 735.5 | 980 | 0.98 | 0.96 |
| 1 Torr | 133 | 1.35×10$^{-3}$ | 1 | 1.33 | 1.33×10$^{-3}$ | 1.31×10$^{-3}$ |
| 1 mbar | 100 | 1.02×10$^{-3}$ | 0.75 | 1 | 10$^{-3}$ | 9.869×10$^{-4}$ |
| 1 bar | 1.0×10$^5$ | 1.02 | 750 | 10$^3$ | 1 | 9.869×10$^{-1}$ |
| 1 atm | 101,325 | 1.03 | 760 | 1,013.25 | 1.01325 | 1 |

The atmospheric pressure exerts a force of 1 kg/cm$^2$ on the vacuum chamber wall. Consequently, vacuum bellows, mechanical systems with asymmetric vacuum vessels dimensions, etc. must be designed to withstand forces up to several hundred kilograms!

In a vacuum system at thermodynamic equilibrium, the particle velocity distribution follows a Maxwell-Boltzmann distribution, the pressure is then given by the ideal gas law, Eq. (1).

$$P = nkT, \qquad (1)$$

with n the gas density in molecules/m³, $k=1.38\times10^{-23}$ J/K the Boltzmann constant, $T$ the gas temperature in K.

The pressure is proportional to the quantity of gas in the vessel and increases linearly with the gas temperature, i.e., with the thermal speed of the molecules. During a bake-out at 300°C, the pressure in a closed vacuum system increases by a factor 2.

In a vacuum system, the gas is usually composed of several types of molecules. The so-called Dalton's law expresses the total pressure, $P_{tot}$ as the sum of all partial pressures, $P_i$, Eq. (2). Air is a mixture of 78% $N_2$, 21 % $O_2$, 1 % Ar, and traces of other gases.

$$P_{tot} = \sum P_i = kT \sum n_i. \qquad (2)$$

When the molecules circulate in a vacuum chamber, they may collide with each other. The mean free path is defined as the average distance a molecules travels between two successive collisions; it is inversely proportional to the pressure.

When evacuating a vacuum chamber from atmospheric pressure down to high vacuum, (HV, $10^{-7}$ mbar) or ultra-high vacuum, (UHV, $10^{-10}$ mbar), the mean free path varies by more than 10 orders of magnitude! In an accelerator, the vacuum system typically operates below $10^{-7}$ mbar, the mean free path is in the kilometre range, much larger than the vacuum chamber dimension. Hence, the molecules interact only with the vacuum chamber walls, this is the molecular regime. In this regime, the vacuum vessel has been evacuated from its volume. The pressure level inside the vessel is dominated by the nature of the surface.

The conductance, $C$, is defined as the ratio of the molecular flux $Q$, to the pressure drop, $P'-P$, along a vacuum vessel, see Eq. (3).

$$C = \frac{Q}{P' - P}, \qquad (3)$$

where $Q$ is the gas flow rate in mbar.ℓ/s and $P'$ (or $P$) the pressure upstream (or downstream) to the flow in mbar.

The conductance is a characteristic of a vacuum vessel. It is a function of the vessel geometry, the nature of the gas and its temperature. It scales like $\sqrt{T/M}$, the higher the molecular mass, the lower the conductance. The unit is usually expressed is ℓ/s.

When two vacuum vessels of conductance $C_1$ and $C_2$ are placed in parallel, their conductances are simply added. However, when vacuum vessels are placed in series, their conductances are added inversely, see Eq. (4).

$$\begin{aligned}\text{In parallel: } & C = C_1 + C_2 \\ \text{In series: } & \frac{1}{C} = \frac{1}{C_1} + \frac{1}{C_2}\end{aligned} \qquad (4)$$

In the molecular regime, for air (M = 29 g/mole) at 20°C (~ 300 K), the engineering formula of the conductance of a (thin) orifice with surface area $A$, and for a tube a tube with diameter $D$ and length $L$, is given by Eq. (5) and by Eq. (6).

$$C_{hole,air,RT}[l/s] = 11.6\, A\, [cm^2]. \tag{5}$$

For air at room temperature, the conductance of a 10 cm diameter orifice equals ~ 900 ℓ/s.

$$C_{tube,air,RT}[l/s] = 12.1\, \frac{D[cm]^3}{L[cm]}. \tag{6}$$

For air at room temperature, the conductance of a 10-m-long tube with a 10 cm diameter, the conductance is 12 ℓ/s. The tube conductance is much lower than the conductance of a hole.

The specific conductance is defined as the conductance of a tube with unit length. For a 10 cm diameter tube, it equals 121 ℓ.m/s.

To increase the conductance of an accelerator vacuum system built with beam tubes, it is better to have a vacuum chamber with a large diameter and short length, as the tube conductance scales with $D^3/L$. Monte Carlo calculations can be used to compute the conductances of complicated geometries.

The gas flow rate is the quantity of gas passing through a tube section per unit time. It is usually expressed in mbar.ℓ/s and is proportional to the number of molecules passing through per second. It has the dimension of J/s, i.e. ,Watt (1 Pa m$^3$ s$^{-1}$ = 1 W). Table 2 gives some usual flow rate units and their conversions to other units.

**Table 2:** Conversion table between usual gas throughput units

| Converts to | Pa m$^3$ s$^{-1}$ | Torr l s$^{-1}$ | mbar l s$^{-1}$ | Molecules s$^{-1}$ (at 300K) |
|---|---|---|---|---|
| 1 Pa m$^3$ s$^{-1}$ | 1 | 7.5 | 10 | 2.46×10$^{20}$ |
| 1 Torr l s$^{-1}$ | 0.133 | 1 | 1.33 | 3.21×10$^{19}$ |
| 1 mbar l s$^{-1}$ | 0.1 | 0.75 | 1 | 2.41×10$^{19}$ |

Imagine that the 10-m-long, 10 cm diameter beam tube of the CERN CMS experiment, coated with a Non-Evaporable Getter (NEG), is exposed to a gas load of 1 mbar.ℓ. Since this film is highly effective at pumping, the surface will be covered by a monolayer of gas (10$^{15}$ molecules/cm$^2$), rendering the film saturated and unusable for subsequent operation. Therefore, 1 mbar.ℓ represent an enormous quantity of molecules, which can be highly detrimental to the vacuum system.

The pumping speed (or volumetric flow rate), *S*, of a pump is defined as the ratio of the flux of molecules pumped, *Q*, to the pressure at the pump inlet, *P*, see Eq. (7). *S* is usually expressed in ℓ/s with *Q* in mbar.ℓ/s and *P* in mbar.

$$S = \frac{Q}{P}. \tag{7}$$

The gas throughput (or gas flow rate), *Q*, in a vacuum pump is the volume of gas, d*V*, moving through the pump inlet per unit of time, *dt*, multiplied by the pressure at the pump, see Eq. (8). When the pressure over the pump increases (assuming constant pumping speed), the gas throughput increases linearly.

$$Q = P\, \frac{dV}{dt} = P\, S. \tag{8}$$

Hence, the pressure in a vacuum system is simply given by the ratio of the gas load to the pumping speed, Eq. (9). Decreasing the gas load (or outgassing) and/or increasing the pumping speed allows for reaching low pressure. Since the range of available pumping speeds is much lower than the range of specific outgassing rates, outgassing must be optimised to achieved UHV.

$$P = \frac{Q}{S}. \tag{9}$$

When placing a vacuum pump of pumping speed *S*, at one end of a vacuum system (e.g., a tube) with conductance *C*, the pressure P' at the other end is a function of *S* and *C*. Using Eq. (3) and Eq. (8), the effective pumping speed, $S_{eff}$, is obtained by identification, as shown in Eq. (10). The effective pumping speed seen at position *P'* is the result of adding the vacuum vessel conductance, *C*, in series with the pump's pumping speed, *S*.

$$\begin{cases} Q = PS \\ Q = C\ (P' - P) \end{cases} \Rightarrow Q = \frac{SC}{S+C}P' = S_{eff}P' \tag{10}$$

Three cases are of particular interest: 1) when *C* = *S* then $S_{eff}$ = *S*/2, 2) when C >> S then $S_{eff}$ = *S* and 3) when C << S then $S_{eff}$ = *C*. In the third case, the system is said to be "conductance limited". Therefore, maximizing the conductance improves the efficiency of the pumping system. A very large conductance is required to fully exploit the benefit of a vacuum pump!

Conductance limitation is a common feature of all accelerators. For lumped pumping systems, it can be shown that increasing the pumping speed much beyond 10-20 times the specific conductance of a beam tube is ineffective unless the distance between pumps is very short (~1 m or less). To overcome this limitation, a distributed pumping system is routinely used in accelerator vacuum systems (HERA, LEP, LHC).

For example, in the cryogenic sections of the LHC, the beam tube provides a relatively low specific conductance (11 ℓ/s.m). The system is conductance limited and relies on the distributed pumping speed provided by the beam screen's holes (360 ℓ/s per meter).

The (intrinsic) specific outgassing rate is the quantity of gas leaving the surface per unit of time and per unit of exposed geometrical surface. The SI unit is $Pa.m^3.s^{-1}.m^{-2}$ (or Pa.m/s or $W/m^2$). Several alternative units are used in the literature. Table 3 provides the conversion factor for common specific outgassing units.

**Table 3:** Conversion table between usual specific outgassing rate units

| Converts to | Pa m $s^{-1}$ | Torr ℓ $s^{-1}$ $cm^{-2}$ | mbar ℓ $s^{-1}$ $cm^{-2}$ | Molecules $s^{-1}$ $cm^{-2}$ |
|---|---|---|---|---|
| 1 Pa m $s^{-1}$ | 1 | $7.5 \times 10^{-4}$ | $10^{-3}$ | $2.46 \times 10^{16}$ |
| 1 Torr ℓ $s^{-1}$ $cm^{-2}$ | $1.33 \times 10^3$ | 1 | 1.33 | $3.27 \times 10^{19}$ |
| 1 mbar ℓ $s^{-1}$ $cm^{-2}$ | $10^3$ | 0.75 | 1 | $2.46 \times 10^{19}$ |

During the manufacturing process of any material, atoms and molecules are sorbed, i.e., adsorbed or absorbed, on the material's surface or within the bulk. Additionally, the surface can be very rough, and the material may be porous. Consequently, the quantities of gas adsorbed and absorbed in any material can be quite large. For illustration, even after baking stainless steel at 400C for 24 hours, ion bombardment in a glow discharge can remove approximately 80 monolayers of CO [3].

The specific outgassing rate, *q*, is therefore an extremely important parameter in vacuum technology. It depends on the nature of the surface, its cleanliness, its temperature, and the pump-down time. As shown in Eq. (9), the final pressure is governed by the outgassing rate, and thus by the specific outgassing rate (*Q* = *A q*).

The specific outgassing rate of metallic surfaces scales as 1/*t* whereas for polymers, it is three orders of magnitude larger and scales as 1/√*t*. A well-designed vacuum system should only use metallic surfaces and avoid polymers (or use them in small quantities for particular applications).

For unbaked metallic surface, H$_2$O is the dominant outgassing species, and the specific outgassing rate is given by Eq. (11).

$$q(t) = \frac{3 \times 10^{-9}}{t[h]} \; mbar \; l \; s^{-1} \; cm^{-2} \tag{11}$$

The pressure in a 10-m-long, 10 cm diameter stainless steel unbaked tube evacuated by a 30 ℓ/s pump reaches 1×10$^{-7}$ mbar (or 7×10$^{-10}$ mbar) after 1 day (or 7 months) of pumping.

An in-situ bake out allows for achieving UHV in a reasonable amount of time. A bake-out above 150˚C allows to remove H$_2$O, after which H$_2$ becomes the dominant gas. A bake-out to higher temperature (e.g., 300˚C for stainless steel) allows for the desorption of molecules with stronger binding energies (e.g., CO), which may later become available for non-thermal desorption induced by circulating beams.

Table 4 provides typical values of specific outgassing rate (in Torr.ℓ/s/cm$^2$) of baked technical surfaces used for the construction of vacuum systems [9]. Values are given after 50 hours of pumping for Al, Cu baked at 150˚C, and stainless-steel at 300˚C.

**Table 4:** Outgassing value in Torr.ℓ/s/cm$^2$ of some baked technical surfaces after 50 h pumping [9].

| Gas | Al | Cu | Stainless steel | Be |
|---|---|---|---|---|
| H$_2$ | 5×10$^{-13}$ | 1×10$^{-12}$ | 5×10$^{-13}$ | 1×10$^{-12}$ |
| CH$_4$ | 5×10$^{-15}$ | 5×10$^{-15}$ | 5×10$^{-15}$ | 1×10$^{-14}$ |
| H$_2$O | 1×10$^{-14}$ | < 1×10$^{-15}$ | 1×10$^{-14}$ | 2×10$^{-14}$ |
| CO | 1×10$^{-14}$ | 1×10$^{-14}$ | 1×10$^{-14}$ | 2×10$^{-14}$ |
| CO$_2$ | 1×10$^{-14}$ | 5×10$^{-15}$ | 1×10$^{-14}$ | 2×10$^{-14}$ |

After bake-out, the pressure in a 10-m-long, 10 cm diameter stainless steel is dominated by H$_2$ and reaches 7×10$^{-10}$ mbar. Successive bakeouts reduce the total outgassing rate by approximately 1.8 per cycle.

Several methods are employed in vacuum technology to reduce the outgassing rates. Chemical cleaning is used to remove gross contamination such as grease, oil, fingerprints [10]. The cleaning process removes the thick, contaminated oxide layer from the vacuum chamber surface and replaces it with a clean oxide layer. Alkaline or acid etching is needed for heavily oxidised surfaces. Passivation is required to form a thin, stable and inert oxide layer. Vacuum firing at 950°C is applied to reduce the hydrogen content for 316LN stainless steel. Glow discharges cleaning is used to remove the adsorbed gases and the metal atoms by sputtering. In all cases, wear clean (and hydrocarbon free!) gloves when handling materials and use clean tools. Dry machining can always be performed using alcohol as a coolant instead of a hydrocarbon-based lubricant. The vacuum components should be designed to be compatible with the various cleaning steps. Care should be taken to avoid small traps where residues from cleaning or etching may accumulate.

At CERN, vacuum firing is routinely used to deplete hydrogen from 316 LN-type stainless steel (316L and 304 series cannot be vacuum-fired due to recrystallization and carbide precipitation at the grain boundaries, which may cause leaks at the flange). The specific outgassing rate of a vacuum-fired 316 LN stainless steel following 300˚C bake out reaches 6×10$^{-15}$ mbar.ℓ/s/cm$^2$. Consequently, the ultimate pressure in a 10-m-long, 10 cm diameter tube evacuated by a 30 ℓ/s pump reaches 6×10$^{-12}$ mbar.

The specific outgassing rate of ceramics depends on their porosity and composition. Al$_2$O$_3$, Macor and nitrides ceramics requires in-situ bakeout at 200˚C to reach approximately 10$^{-11}$ mbar.ℓ/s/cm$^2$ at room temperature. Ferrites, such as TT2-111R, CMD5005 and CMD10, are treated at temperatures between 400°C and 1000°C to achieve a specific outgassing rate of approximately $10^{-11}$ mbar.ℓ/s/cm$^2$ after bake out. But by design, these materials can heat up during beam operation.

Polymers have a large permeability to helium, so care must be taken during leak detection. They also have a large solubility for water, so they should be protected from humidity exposures. Dry air or nitrogen should be used for venting. Polymers can be baked to 150-200 °C. Thick polymers materials require long pumping time due to diffusion effects. For example, unbaked elastomers such as Viton yields $2\times10^{-7}$ mbar.ℓ/s/cm$^2$ after 10 hours of pumping and reach $5\times10^{-10}$ mbar.ℓ/s/cm2 after bake-out. Hard plastics like 0.1 mm thick unbaked Kapton yields $5\times10^{-10}$ mbar.ℓ/s/cm$^2$ after 100 hours of pumping.

Material with low vapour pressure should always be used for the design. Zinc, cadmium, lead, sulphur, and phosphorous are forbidden in vacuum systems. Reminder: brass is made of Cu+Zn! During the bake-out of brass, Zn evaporates and coats the inner part of the vacuum vessel.

Common welding process used in vacuum technology includes TIG, electron beam and laser welding; vacuum brazing is also used for specific applications [11, 12]. Some guidelines for UHV application include: use protective gas for TIG; butt welds should be performed on the vacuum side or be fully penetrant; fillet welds should not be fully penetrant; welding on the inside and outside of the chamber makes the leak detection impossible. If mechanical strength requires it, a continuous weld on the vacuum side can be combined with stitch welds on the opposite side. In cryogenic machines, full penetrating welds between the vacuum and helium enclosures are forbidden; seamless beam tubes must be used. Soldering flux remover is forbidden because its residues are difficult to remove and can cause leaks. Proper weld preparation is also crucial to ensure weld quality and avoid contamination.

## 2  Vacuum components

Pressure monitoring in an accelerator is typically performed using Pirani and Penning gauges. Pirani gauges are used in the range 1 atm to $10^{-4}$ mbar to monitor the roughing of the vacuum system. These gauges are robust but gas dependent, with an accuracy range of 10-100 %. The pressure measurement is based on the variation in heat conductivity with the gas pressure, meaning pressure readings above 1 mbar and below $10^{-3}$ mbar are inaccurate.

Penning gauges are commonly used in the range $10^{-5}$ to $10^{-10}$ mbar, typically for interlocking purposes. These gauges are also robust, gas dependent, and have an accuracy range of 20-50 %. A Penning gauge is a cold-cathode ionisation gauge, meaning it does not use a hot filament; electrons are produced via field emission. The operating principle is based on measuring a discharge current in a Penning cell, which is a function of pressure: $I^+ = P^n$, with $n$ close to 1. When operated at pressure higher than a few $10^{-6}$ mbar, the discharge becomes unstable, and materials can be sputtered during extended operation, potentially leading to false or unstable pressure readings. At low pressure, the discharge may extinguish, resulting in a zero pressure reading.

Bayard-Alpert gauges are used for vacuum measurement purposes in the range $10^{-5}$ to $10^{-12}$ mbar. These are hot-cathode gauges, where electrons emitted by the hot filament oscillate inside the grid and ionise the molecules of the residual gas. Ions are then collected by an electrode. The gauge requires calibration. In UHV, the ion current typically ranges from 0.1 to 10 pA, so the collecting cable must be shielded from electromagnetic perturbations. For UHV measurements, the filament must be degassed. The typical outgassing rate is 5 to $10\times10^{-10}$ mbar.ℓ/s and can be further reduced. When using a thin ion collector, the X-ray limit due to photoelectrons leaving the ion collector is approximately $2\times10^{-12}$ mbar.

Residual Gas Analysers (RGAs) are used in the range $10^{-4}$ to $10^{-12}$ mbar and are primarily used for gas analysis. A hot filament produces electrons that ionise the residual gas inside a grid. A mass filter is introduced between the grid and the ion collector. The ion current can be measured in Faraday mode or in secondary electron multiplier mode to improve the detection limit. RGAs are expensive and delicate instruments that produces spectra than can sometimes be difficult to analyse. They can also be used to identify leaks by using Ar and $N_2$ as tracers (masses 40, 14 and 28). The instrument needs to be properly tuned and calibrated, and its filament must be degassed. Its outgassing ranges from $2\times10^{-9}$ to $5\times10^{-7}$ mbar.ℓ/s.

In the accelerator tunnel, turbomolecular pumping groups are used to pump down from atmospheric pressure and to commission vacuum sectors down to $10^{-11}$ mbar. These systems are mobile and consist of a rotary vane primary pump combined with a turbomolecular pump. There are two categories of primary pump: dry and wet.

Dry pumps are expensive and may need additional cooling, such as water. Wet pumps, on the other hand, operate with oil, which serves as a sealant, lubricant, heat exchanger, and protection against rust and corrosion. A filter for oil vapor is required. The end pressure lies between $10^{-3}$ and $10^{-2}$ mbar, and the pumping speed ranges from 4 to 40 $m^3$/h.

The turbomolecular pump operates in the molecular regime, with a constant pumping speed below $10^{-3}$ mbar, typically ranging from 10 to 3 000 ℓ/s. The pumping speed is proportional to the area of the entrance flange times the rotational speed of the blades. The pumping mechanism is based on the transfer of momentum from the rotating blades (operating at speed up to 40 000 turns/min) to the gas molecules, directing them towards the exhaust. Each rotor-stator stage contributes incrementally to this process, a 60 ℓ/s turbomolecular pump typically contains around 10 stages. Thanks to its high compression ratio for heavy gases ($10^3$ for $H_2$ and $10^9$ for $N_2$), turbomolecular pumps enables a "clean" vacuum with minimal hydrocarbons contamination. When baked, the ultimate pressure can reach a few $10^{-11}$ mbar.

The sputter ion pump operates in the range $10^{-5}$ to $10^{-11}$ mbar and is widely used to maintain the pressure in the vacuum chamber of an accelerator. It is a capture pump with no exhaust. Its pumping speed ranges from 1 to 500 ℓ/s, and it decreases as the pressure lowers. The pump consists of several Penning cells. Electrons spiral along the magnetic field lines within the Penning cell and ionised the residual gas molecules. The ions are then accelerated towards the cathode (at few kV), where they sputter titanium (Ti). Titanium forms a chemical bond with gas molecules, trapping them. Hydrogen diffuses into the Ti cathode and hydrocarbons are cracked into C, H and O, which can be chemisorbed. Noble gases, which does not react with Ti, are buried or implanted into the cathode. Similar to the Penning gauge, the ion current reading is proportional to the pressure. This pump is also used as an interlock.

Getters (e.g. titanium) are materials capable of chemically adsorbing gas molecules. For a getter to be effective, its surface must be clean. A Ti sublimation pump is an evaporable getter pump. For Non-Evaporable Getters, such as the TiZrV films used in the room temperature parts of the LHC, a clean surface is achieved by heating the film to 200°C to dissolve the native oxide layer into the bulk material. NEGs can pump most gases at room temperature, except for rare gases and methane, so they are often combined with ion pumps. The 1 μm thick TiZrV film is typically deposit by magnetron sputtering at 300°C using Kr gas. These films exhibit very high pumping speed: 250 ℓ/s/m for $H_2$ and 20 000 ℓ/s/m for CO. Therefore, they provide a distributed pumping. They have also a very low outgassing rate ($10^{-18}$ mbar.ℓ/s for methane) but they have a limited pumping capacity, are fragile, and are sensitive to pollutants such as hydrocarbons and fluorine. Photons and electrons induced desorption yields of TiZrV film are also extremely low.

For medium and high vacuum systems, elastomer seals (Viton®, Nitrile, EPDM) and clamp flanges are typically used. Many fittings – such as elbows, bellows, T, cross, flanges with short pipe, reductions, blank flanges- are available with several ISO diameters and commonly use KF-type clamps, though other types of clamps are also available.

For ultra-high vacuum systems, metallic gaskets and bolted flanges are standard. Several components utilise Conflat® type technology, including blank flanges, rotatable flanges, welding flanges, elbows, Ts, crosses, adaptors, zero length double-side flanges, and windows. Copper gaskets (and sometime aluminum gaskets) are available in various ISO diameters. OFE (Oxygen Free Electronic) / OFHC (Oxygen Free High Conductivity) Cu is standard but silver-coated OFS (Oxygen Free Silver-bearing) Cu must be used for bake-out. A4-100 screws and Mo or Ag-coated nuts are recommended for assembly. Collars chains can also be used when a quick connection is needed, for example in a radiation environment.

For vacuum system construction, metallic tubes are preferred to minimise outgassing. Stainless steel is commonly chosen for it mechanical properties (such as ease of machining and welding), though vacuum chambers can also be made from aluminum, copper, or beryllium [13, 14]. The material shall be low cost, temperature-resistant (to withstand bake-out), corrosion-resistant, and exhibit low induced radioactivity, high mechanical strength, as well as good thermal and electrical conductivity. Bellows should be equipped with RF bridges to reduce beam impedance. Right-angle valves are used for roughing, while gate valves (often RF-screened) are employed for vacuum sectorisation.

## 3      Getting ready with beams: from construction to operation

Vacuum is a common feature of all the components of an accelerator. By design, each individual part must meet strict vacuum specifications, including outgassing, gas composition, and ultimate pressure. Depending on the accelerator's operating parameters, additional properties, such as bakeability, thermal and electrical conductivity, machinability, and cost may also be required.

Given the complexity of an accelerator, all equipment within vacuum undergoes vacuum acceptance tests prior to installation. These tests are crucial for ensuring the accelerator's availability. Standard tests include functional checks, pump-down, leak detection, residual gas composition analysis, and total outgassing rate measurements. For certain systems, a bake-out may also be necessary. These acceptance tests can take several weeks for some components, making careful production planning essential.

The construction of an accelerator also involves specific integration studies to evaluate the space required by the equipment in the tunnel, the feasibility of installation and removal, and accessibility for maintenance and repair. For large machines, the use of a database is essential for managing efficiently the large set of components.

The accelerator vacuum system is divided into discrete vacuum sectors, which are isolated from each other by gate valves. For example, the LHC contains approximately 300 gate valves. Each vacuum sector is equipped with two gate valve assemblies, one at each extremity, and a measurement and pumping port at the centre of the sector. Gate valves are interlocked to close automatically if the pressure in their vicinity exceed a preset threshold, triggering a beam dump. The closure of a gate valve by an interlock prompts the closure of both upstream and downstream gate valves, effectively isolating the affected vacuum sectors. During maintenance periods, gate valves are closed to safeguard the entire vacuum system in case of an incident and to allow venting and recommissioning when repairs are required in a specific vacuum sector. Atmospheric venting must be carried out using dry nitrogen to avoid reaction between oxygen or water and carbon, which is present on the vacuum chamber surface in the form of graphite. This prevents a subsequent increase in gas desorption coefficients [3].

Each vacuum sector contains multiple vacuum chambers, which connected by compensator bellows. These bellows are essential for accommodating thermal expansion during bake-out and for handling vacuum forces caused by differential pressure. To prevent damage and ensure proper operation, each vacuum chamber must have a single fixed point, while any additional support must be sliding. The compensator bellows are equipped with RF bridge to minimise impedance and ensure the system's performance during beam operation.

Installation drawings are crucial for the proper assembly of an accelerator machine. These drawings specify, for each vacuum sector, the positions and lengths of all components, including vacuum equipment. Components are named following a standardised convention: for instance, all vacuum equipment begins with the letter "V", and "VC" stands for vacuum chamber. These drawings are automatically generated from the equipment database to ensure consistency.

Before installation, the survey team marks the floor to indicate the precise location where the supports will be placed. The marked positions are drilled, and the supports are installed and pre-aligned. After vacuum acceptance tests are completed, sector valve assemblies are positioned and aligned. Following this, the vacuum chambers are placed on the supports and aligned with high precision. Typically, an alignment accuracy of ±1–2 mm can be easily achieved using simple tooling. The installation of the first gate valve assemblies and vacuum chambers in the LHC occurred between June and July 2006, during the construction of vacuum sector A5L8.

Once the alignment is complete, the vacuum sector is closed by inserting the vacuum modules, and the system is sealed by tightening the bolted flanges. At this point, the pump-down process and leak detection procedure can commence. Any leaks greater than $10^{-11}$ mbar.ℓ/s must be identified and sealed.

During pump-down, early signs of leak can be identified. **Fig. 1** illustrates the pump-down of an unbaked 10-m-long 8 cm diameter beam pipe. In the absence of large leaks, the pressure reduction typically follows a $1/t$ slope, where t is the pumping time. However, if there are leaks greater than $10^{-6}$ mbar.ℓ/s, a deviation from the expected slope becomes evident after several hours of pumping.

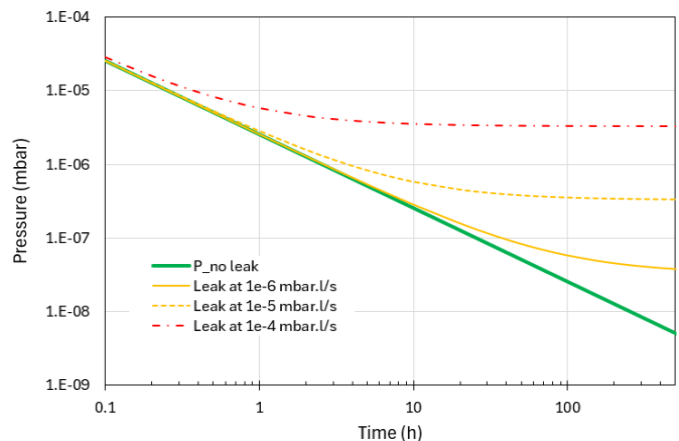

**Fig. 1:** Schematic of a Bayard-Alpert gauge.

During leak detection, helium is sprayed around the test piece, and a leak detector (or an RGA tuned to the He signal) is connected to the device under test. Common leak sources include gaskets (due to poor installation or inadequate thermalization during bakeout), feedthroughs, welds, corrosion points, beam-induced heating points, and material porosities. However, thanks to the vacuum acceptance tests performed on each sub-component before installation, the global leak check should primarily focus on the joints made during the final installation phase in the accelerator.

A leak detector is a specialised instrument designed to detect and measure leaks. In the leak detection cell, the residual gas is ionized, and the resulting ions are accelerated and deflected by a magnetic sector for detection. The magnetic sector separates helium from other gases. The leak detection cell is pumped by a turbomolecular pump, which enable helium detection using the counter-flow method. This method leverages the high compression ratio of the turbomolecular pump for heavy gases, enriching the signal inside the leak detection cell with lighter gases. A second turbomolecular pump inside the apparatus ensures clean pumping of the device under test. In-situ calibration of the leak detector is possible, and when properly used, the detection limit is $10^{-12}$ mbar.ℓ/s ($10^{-13}$ Pa.m$^3$/s).

Virtual leaks originate from the leakage of atmospheric air from trapped volumes. They must be eliminated during the design phase. For instance, care must be taken to provide ventilation holes for screws placed in blind holes. Argon, used for inerting during welding, may also leak from porous welds. Atmospheric air may diffuse out of a porous wall (or, in extreme cases, through the wall itself). X-ray inspection of the material is used for quality control. Three-dimensionally forged flange material must be used (rolled material should not be used).

Once the vacuum sector is mounted, aligned, pumped down, and leak-tight, the bake-out phase can start. All components must be baked together, including the beam pipes inside the magnets. Cold spots are not allowed. It is better to bake at a higher temperature than to extend the bake-out duration (due to hydrogen diffusion and the increased risk of power cuts during longer bake-out periods). A bake-out above 130–150˚C removes water.

Beware of potentially long-time constants and low-temperature reach for components placed inside a vacuum system (due to poor thermal conductivity or heavy objects). Appropriate mechanical design is required!

The bake-out is performed using a Programmable Logic Controller (PLC). The PLC controls the heating cycle, provides interlocks, and logs the data. Typical materials used for bake-out include collars of different diameters and lengths; thermocouples, usually type E (Chromel-Constantan); bake-out jackets that can precisely fit any component; heating tapes ranging from 100 W to 2 kW, insulation made of glass/mineral fiber, polymer foam, or aerogel and rack controllers

Remember: all parts must be baked, including components inside vacuum devices, as cold spots are forbidden.

In the LHC, the typical commissioning of a room-temperature vacuum sector is performed in two steps. During the first step, all stainless-steel components (collimators, beam instruments, etc.) are baked at 300˚C for 24 hours, while the vacuum chambers coated with TiZrV are held at 100˚C. In the second step, the temperature of the stainless-steel parts is decreased to 120˚C, and the various instruments are degassed. Then, the temperature of the NEG pipes is increased to 200˚C for 24 hours for activation. The entire process lasts about a week, including the final qualification step, pinch-off, and removal of the mobile instruments. As a result of this procedure, the average pressure in the room temperature vacuum sector was in the range of $10^{-11}$ mbar at the LHC startup [15].

While the vacuum system has been properly designed, prototyped, produced, qualified, assembled, and commissioned, one important factor remains: the circulation of beams. All modern accelerators are highly demanding in terms of performance. Intense and bright beams require proper management of beam impedance. If this is not achieved, the system is exposed to unwanted beam-induced heating, outgassing, and potential damage to fragile components. The LHC is no exception [16]. High-energy synchrotrons must contend with synchrotron radiation; this applies to all electron machines, and the LHC was the first to experience this effect with protons. The emitted light is a result of the bending of charged particles in the machine's arcs. The energy can be large enough to melt components if not properly managed.

From a vacuum perspective, photon irradiation stimulates molecular desorption. During continuous beam operation, the desorption yields reduce from approximately $10^{-2}$ to $10^{-7}$ molecules/photon for a conditioned surface. However, due to the large number of photons produced, the resulting pressure can be much higher than the original vacuum pressure without beam. Consequently, in most accelerator machines, this "dynamic" pressure becomes dominant over the "static" pressure.

The circulating beam may also trigger the build-up of an electron cloud. In the LHC, the proton bunches are closely spaced, separated by 25 ns (7.5 m). The photoelectrons, originating from synchrotron radiation, are accelerated to about 100 eV by the dense proton bunch ($1.1 \times 10^{11}$ protons/bunch). This leads to the emission of a cascade of secondary electrons. The phenomenon is primarily driven by the secondary electron yield (SEY), a material property. These electrons also provoke molecular desorption, which adds to the dynamic pressure induced by synchrotron radiation. In cryogenic accelerators, the electron cloud-induced heat load can also be significant, as in the case for the LHC.

Usual materials have a SEY in the range of 2–3. The solution to the electron cloud build-up issue in the LHC arcs is beam scrubbing, which causes graphitisation of the surface and reduces SEY to close to 1. In the LHC room-temperature vacuum system, including the experiments, the NEG coating provides an SEY of ~1.1, preventing multipacting. For the HL-LHC, the new inner triplet and machine sections will be coated with amorphous carbon, which has an SEY of unity.

Pressure runaway due energetic ion bombardment of the vacuum chamber wall is another limiting phenomenon that must be considered during the design phase. Additional details on vacuum systems for accelerators can be found in [17].

## 4     Summary

The ideal gas law, Dalton's law, and the Maxwell-Boltzmann distribution are used to describe gas kinetics in a vacuum system. Due to the typically large mean free path, most vacuum systems operate in the molecular flow regime. A vacuum system can be analysed using the concepts of conductance, pumping speed, and outgassing, although detailed calculations can be performed with specialized computational tools. A wide range of instruments, materials, techniques, technologies, methods, and empirical data is available to support the design, fabrication, and operation of accelerator vacuum systems. This systematic approach was instrumental in the development and successful operation of the LHC vacuum system [18, 19, 20].